\documentclass[amsmath,amssymb,floatfix]{epl}
\usepackage{graphicx}
\usepackage{bm}

\title{Upper critical field $H_{c2}$ in Bechgaard salts \chem{(TMTSF)_2 PF _6}}
\author{Ana Dom\'{\i}nguez Folgueras\inst{1,2,3} \and Kazumi Maki\inst{3}}
\institute{
  \inst{1} Departamento de F\'{\i}sica, Universidad de Oviedo, 33007 Oviedo, Spain\\
  \inst{2} Instituto de Ciencia de Materiales de Madrid, CSIC, Cantoblanco, 28049 Madrid, Spain \\
  \inst{3} Department of Physics \& Astronomy, University of Southern California, Los Angeles, CA 90089-0484, USA
}
\pacs{74.70.Kn}{Organic Superconductors}
\pacs{74.20.Rp}{Pairing symmetries}
\pacs{74.25.Op}{Mixed states, critical fields, and surface sheaths}

\begin{document}

\maketitle

\begin{abstract}
The symmetry of the superconducting order parameter in Bechgaard salts is still
unknown, though the triplet pairing is well established by NMR data and large
upper critical field 
$H_{c2}(0)\sim5$ Tesla for $\vec{H}\parallel a$ and $\vec{H}\parallel b^\prime$.
Here we examine the upper critical field of a few candidate superconductors 
within the standard formalism. The present analysis suggests strongly chiral 
f-wave superconductor somewhat similar to the one in Sr$_2$RuO$_4$ is the most
likely candidate.
\end{abstract}

\section{Introduction}

The Bechgaard salts (TMTSF)$_2$PF$_6$ is the first organic superconductor 
discovered in 1980~\cite{Jer}. For a long time the superconductivity is
believed to be conventional s-wave~\cite{Ish}. Recently the symmetry of the
superconducting energy gap becomes the central issue~\cite{Sig,KSD}. The upper
critical field at $T=0K$, $H_{c2}\sim 5$ Tesla for both $\vec{H}\parallel a$ and 
$\vec{H}\parallel b^\prime$ for both (TMTSF)$_2$PF$_6$~\cite{Lee} and (TMTSF)$_2$ClO$_4$~\cite{OhN} are clearly beyond the Pauli limit~\cite{Clo,Cha} indicating the 
triplet pairing. More recently the NMR data from (TMTSF)$_2$PF$_6$~\cite{LBC} indicates clearly
the triplet pairing. Therefore the candidates for the superconductivity in 
Bechgaard salts are more likely within p-wave and f-wave superconductors. In
the following we shall examine the upper critical field of these superconductors
following the standard method initiated by Gor'kov~\cite{Gor} and extended by
Luk'yanchuk and Mineev~\cite{Luk} for unconventional superconductors. 
Also we take the quasiparticle energy in the normal state as in the standard
model for Bechgaard salts~\cite{Ish}
\begin{equation}
\xi(k)=v(|k_a|-k_F)-2t_b\cos{bák}-2t_c\cos{cák}
\end{equation}
with $v:v_b:v_c\sim1:1/10:1/300$ and $v_b=\sqrt{2}t_b b$ and $v_c = \sqrt{2}t_c c$. There are
earlier analysis of $H_{c2}$ of Bechgaard salts starting from the one
dimensional models~\cite{Leb,Dep}. However, those models predict diverging 
$H_{c2}(T)$ for $T\rightarrow 0 K$ or the reentrance behaviour, which have not been observed
in the experiments~\cite{Lee,OhN}. Also, the quasilinear T dependence of
$H_{c2}(T)$ in both (TMTSF)$_2$PF$_6$ and (TMTSF)$_2$ClO$_4$ is very unusual. 
Among the models we have considered, the chiral f$^\prime$-wave superconductor with
$\Delta\left(\vec{k}\right)\sim \left(\frac{1}{\sqrt{2}} sgn\left(k_a\right)+\imath\,\sin{\chi_2}\right)\cos{\chi_2}$,
looks most promising, where $\chi_1=\vec{b}\,\vec{k}$ and $\chi_2=\vec{c}\,\vec{k}$
where $\vec{b}$ and $\vec{c}$ are crystal vectors. 

Also if the superconductor belongs to one of the nodal superconductors~\cite{Won}
and if nodes lay parallel to $\vec{k_c}$ within the two sheets of the Fermi surface,
the angle dependent nuclear spin relaxation rate $T_1^{-1}$ in a magnetic field
rotated within the $b^\prime-c^*$ plane will tell the nodal directions.

\begin{figure}
\begin{center}
\includegraphics[width=6cm,clip]{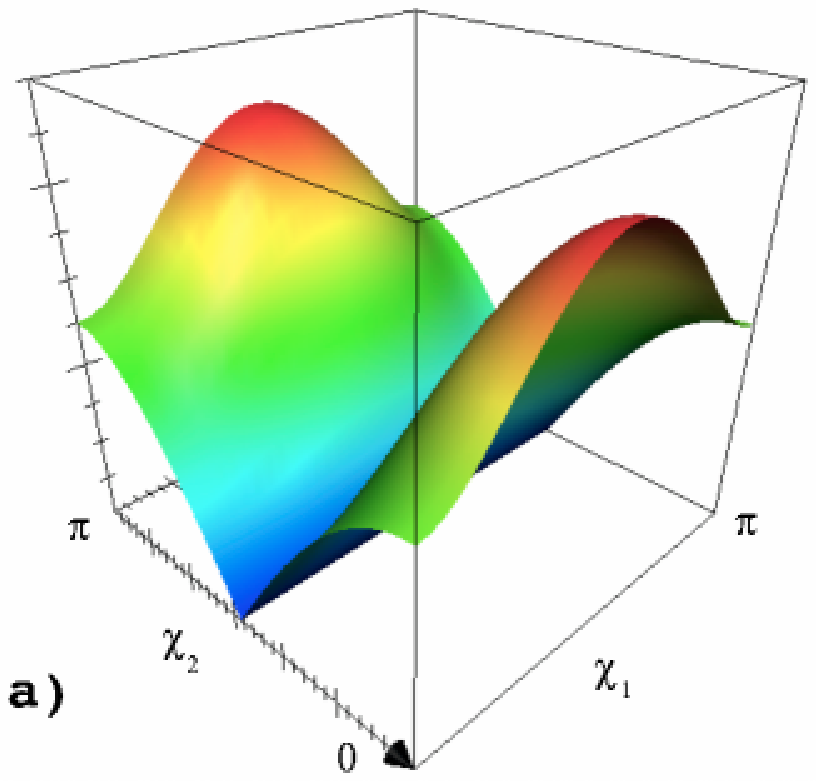}
\includegraphics[width=6cm,clip]{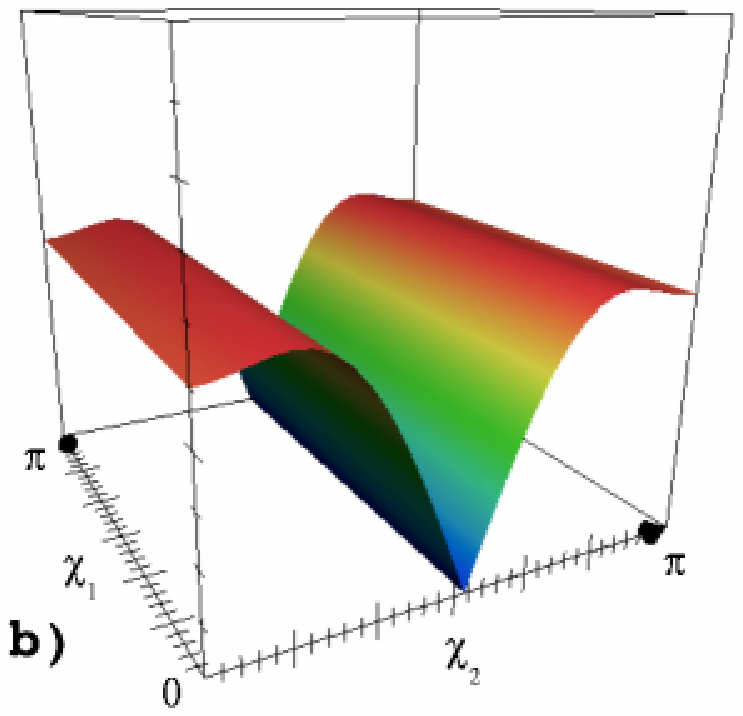}
\caption{$|\Delta(\vec{k})|$ of chiral f-wave and chiral f$^\prime$-wave SC are sketched in a) and b) respectively.}
\label{gaps}
\end{center}
\end{figure}

Before proceeding, we show $|\Delta(\vec{k})|$ os two chiral f-wave superconductors in Fig.~\ref{gaps} a) and b).
where $|\Delta(k)|\sim \left[ (1+\cos{2\chi_1})(1-\frac{1}{2}\cos{2\chi_2})\right]^{\frac{1}{2}}$  and 
$|\Delta(k)|\sim \left[ (1+\cos{2\chi_2})(1-\frac{1}{2}\cos{2\chi_2})\right]^{\frac{1}{2}}$ for chiral f$_1$ and
chiral f$_2$ respectively.

\section{Upper critical field for $\vec{H}\parallel \vec{b^\prime}$}

In the following we neglect the spin component of $\vec{\Delta}(\vec{k})$. Most likely the equal spin pairing is realised in Bechgaard salts as in Sr$_2$RuO$_4$~\cite{KSD}. In this case the spin component
is characterised by a unit vector $\hat{d}$. Also $\hat{d}$ is most likely oriented parallel to $\vec{c^*}$.
Let's assume 
$\hat{d}\parallel\vec{c^*}$, though $H_{c2}(T)$ is independent of $\hat{d}$ as long as the spin orbit
interaction is negligible. Experimental data from both UPt$_3$ and Sr$_2$RuO$_4$ indicate that the 
spin-orbit interactions in these systems are not negligible but extremely small~\cite{Mhp}.
We consider a variety of triplet superconductors (see Fig.~\ref{gaps}):

\subsection{A. Simple p-wave SC: $\vec{\Delta}(k)\sim sgn(k_a)$}
\label{sec_pw}

Following~\cite{WoM} the upper critical field is determined by

\begin{eqnarray}
-\ln{t}=\int_0^\infty \frac{du}{\sinh{u}}\left(1-K_1\right) \\
-C\,\ln{t}=\int_0^\infty \frac{du}{\sinh{u}}\left(C-K_2\right)
\label{main_eq}
\end{eqnarray}

where 
\begin{center}
\begin{eqnarray}
K_1=\langle e^{-\rho u^2 |s|^2}\left(1+2\,C\,\rho^2 u^4 s^4\right)\rangle  \\
K_2=\langle e^{-\rho u^2 |s|^2}\left(\frac{1}{6}\rho^2u^4s^{*4}+C\left(1-8\rho u^2|s|^2+12\,\rho^2u^4|s|^4-\frac{16}{3}\rho^3u^6|s|^6+\frac{2}{3} \rho^4u^8|s|^8\right)\right) \rangle
\end{eqnarray}
\end{center}
and $t=\frac{T}{T_c}$, $\rho = \frac{v_a v_b e H_{c2}(T)}{2(2\pi T)^2}$, 
$s=\frac{1}{\sqrt{2}}sgn(k_a)+\imath \sin{\chi_2}$, $\chi_2=\vec{c}\vec{k}$ and $\langle \ldots \rangle $
means average over $\chi_2$. Here $v_a$, $v_c$ are the Fermi velocities parallel to the a axis and 
the c axis respectively.

Here we assumed that $\Delta(\vec{r})$ is given by~\cite{WoM}

\begin{equation}
\Delta(\vec{r}) \sim \left(1+C(a^+)^4\right)\,\rangle
\end{equation}

where $\rangle = \sum C_n\, e^{-eBx^2-nk(x+\imath z)-\frac{(nk)^2}{4eB}}$ is the Abrikosov 
state~\cite{Abr} and $a^+=\frac{1}{\sqrt{2eB}}\left(-\imath\partial_z - \partial_x + 2\imath eHz\right)$ is the raising operator.

Then in the vicinity of $t\rightarrow1$ we find $\rho=\frac{2}{7\zeta(3)}(-\ln{t})=0.237697(-\ln{t})$ and $C=-\frac{93\zeta(5)}{647\zeta(3)}\rho$.

For $t\rightarrow0$ on the other hand we obtain

\begin{equation}
\rho_0=\lim_{t\rightarrow0}\rho\,t^2=\frac{v_a v_c e H_{c2}(0)}{2\left(2\pi T_c\right)^2}=0.1583
\end{equation}

and $C=-0.031$. From these we obtain

\begin{equation}
h(0)=\frac{H_{c2}(0)}{\frac{\partial H_{c2}(t)}{\partial t}\vert_{t=1}}=0.6659
\end{equation}

\begin{figure}
\begin{center}
\includegraphics[width=6.5cm,height=6cm]{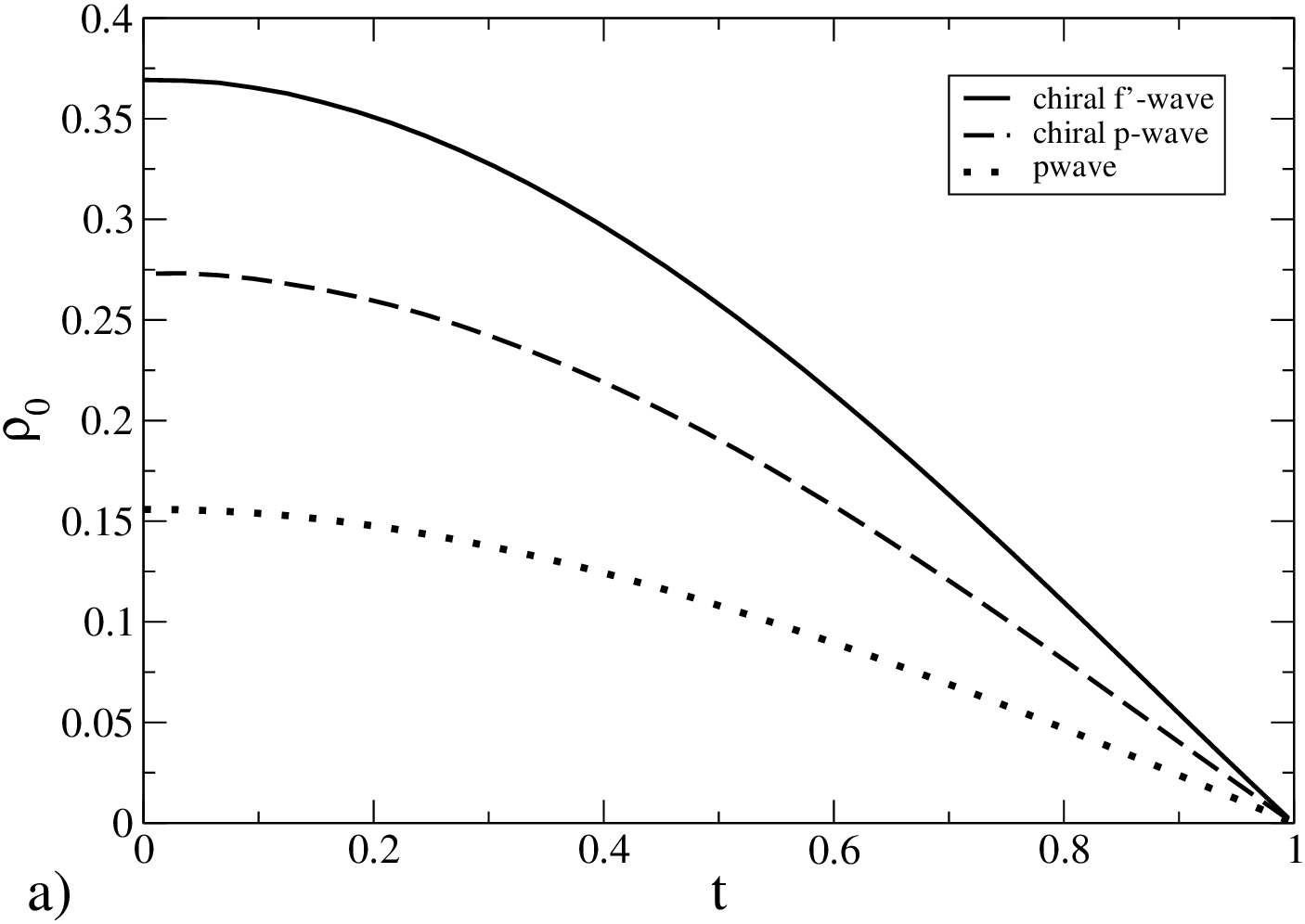}
\includegraphics[width=6.5cm,height=6cm]{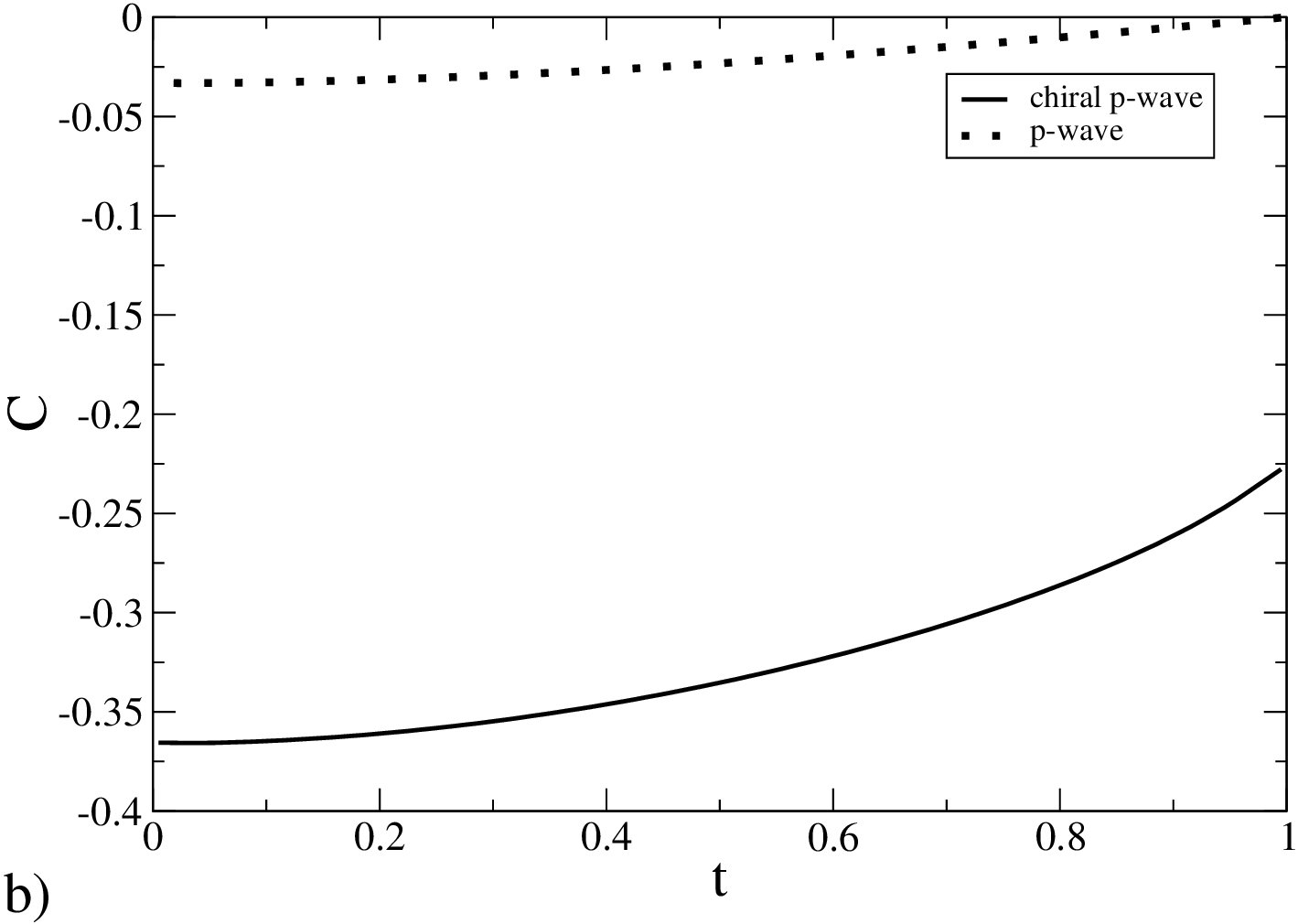}
\caption{Normalised $H_{c2}(t)$ and $C(t)$ for $\vec{H} \parallel \vec{b}^\prime$ are shown in a) and b) respectively. Here solid, dashed and dotted lines are chiral f$^\prime$-wave, chiral p-wave and simple p-wave respectively.}
\label{campob}
\end{center}
\end{figure}

Both $\rho_0(t)$ and $C(t)$ are evaluated numerically and shown in Fig.~\ref{campob} a) and b)
respectively. Here $\rho_0(t)=t^2\rho(t)=v\,v_c e H_{c2}(t)/2(2\pi T_c)^2$.

\subsection{B. Chiral p-wave SC: $\vec{\Delta}(k)=1/\sqrt{2}sgn(k_a)+i\sin(\chi_2)$}
\label{sec_cp}

Here $\frac{1}{\sqrt{2}}sgn(k_a)+i\sin(\chi_2)$ is the analogue of $e^{\imath \phi}$ in the 3D systems in the quasi 1D system.
For a chiral state the Abrikosov function is written as~\cite{Wan}

\begin{equation}
\Delta(\vec{r},\vec{k})\sim(s+Cs^*(a^\dag)^2)\,\rangle
\end{equation}

where $s=\frac{1}{\sqrt{2}}sgn(k_a) + \imath \sin(\chi_2)$. Then we obtained eq.~\ref{main_eq} with

\begin{eqnarray}
K_1=\langle e^{-\rho u^2 |s|^2}\left(|s|^2-2\,C\,|s|^{4}\right)\rangle \\
K_2=\langle e^{-\rho u^2 |s|^2}\left(-|s|^4+C\,|s|^2\left(1-4\rho u^2|s|^2+2\,\rho^2u^4|s|^4\right)\right)\,\rangle
\end{eqnarray}

and the same expressions for $t$, $\rho$,\ldots

For $t\rightarrow1$ we find $C=1-\sqrt{1.5}=-0.2247$ and
$\rho=0.3838 (-\ln{t})$.

On the other hand, for $t\rightarrow 0$ we obtain $C=-0.3660$ and $\rho_0=0.27343$.

From these we obtain $h(0)=0.71324$. We obtain $\rho(t)$ and $C(t)$ numerically.
They are shown in Fig.~\ref{campob} a) and b) respectively.

\subsection{C. Chiral f-wave SC: $\hat{\Delta}(k)\sim \hat{d}s\cos{\chi_1}$}
\label{sec_cf1}

$H_{c2}(t)$ is determined from eq.~\ref{main_eq} where now:
\begin{eqnarray}
K_1=\langle \left(1+\cos{2\chi_1}\right)e^{-\rho u^2 |s|^2}\left(|s|^2-2\rho u^2 |s|^4\right) \rangle \\
K_2=\langle \left(1+\cos{2\chi_1}\right)e^{-\rho u^2 |s|^2}\left(-\rho u^2 |s|^4+C|s|^2\left(1-4\rho u^2|s|^2+2\rho^2u^4|s|^4\right)\right)\rangle
\label{cfw}
\end{eqnarray}

Here now $\langle \ldots \rangle$ means the average over both $\chi_1$ and $\chi_2$. Then it is
easy to see that the chiral f-wave SC has the same $H_{c2}(t)$ and $C(t)$ as the chiral p-wave SC, since the variable $\chi_1$ is readily integrated out.

\subsection{D. Chiral f$^\prime$-wave SC: $\hat{\Delta}(k)\sim\hat{d}s\cos{\chi_2}$}
\label{sec_cf2}

Now we have a set of equations similar to the chiral f-wave except $1+\cos{2\chi_1}$ in both 
eqs.~\ref{cfw} has to be replaced by $\frac{4}{3}\left(1+\cos{2\chi_1}\right)$. Then we obtain for 
$t \rightarrow 1$ $C=-0.2247$ and $\rho=0.5181 (-\ln{t})$. On the other hand, for $t\rightarrow 0$ we find $C=-0.3660$ and $\rho_0=0.3734$.

We show $\rho_0$ and $C(t)$ of the chiral f$^\prime$-wave in Fig.\ref{campob} a) and b)
respectively.

Note that $C(t)$ is the same for three chiral states (chiral p-wave, chiral f-wave and chiral 
f$^\prime$-wave) as well as chiral p-wave studied in~\cite{Wan}

Therefore for the magnetic field $\vec{H}\parallel b^\prime$, the chiral
f$^\prime$-wave have the largest $H_{c2}(t)$ if we assume $T_c$ and $v$, $v_c$ are the same. Also
$H_{c2}(t)$ of these states are closest to the observation.

\section{Upper critical field for $\vec{H}\parallel \vec{a}$}

\subsection{A. Simple p-wave SC: $\Delta\left(\vec{k}\right)=sgn\left(k_a\right)$}

The equation for $H_{c2}(t)$ is given by~\cite{WoM} and can be written as in eq.\ref{main_eq}
with

\begin{eqnarray}
K_1=\langle e^{-\rho u^2 |s|^2}\left(1+2C\rho^2 u^4 |s|^4\right) \rangle \\
K_2=\langle e^{-\rho u^2 |s|^2}\left(\rho^2 u^4 |s|^{4}+C\left(1-8\rho u^2|s|^2+12\rho^2u^4|s|^4-\frac{16}{3}\rho^3u^6|s|^6+\frac{2}{3}\rho^4u^8|s|^8\right)\right)\rangle
\end{eqnarray}

where $t=\frac{T}{T_c}$, $\rho=\frac{v_av_be H_{c2}(t)}{2(2\pi T)^2}$ 
and $s=\sin{\chi_1}+\imath \sin{\chi_2}$ with $\chi_1=\vec{b}\vec{k}$ and 
$\chi_2=\vec{c}\vec{k}$.

Then for $t\rightarrow1$, we find
$C=-\frac{93\zeta(5)}{508\zeta(3)}\rho$
and $\rho=\frac{2}{7\zeta(3)}(-\ln{t})=0.2377(-\ln{t})$.
While for $t\rightarrow0$ $C=\frac{3}{2\beta_0}-\sqrt{\left(\frac{3}{2\beta_0}\right)^2+\frac{1}{12}}=-0.0170129$ and $\rho_0=\frac{v_a v_b e H_{c2}(0)}{2(2\pi T_c)^2}=\frac{1}{4\gamma}\exp{\left[\alpha_0+2C\beta_0\right]}=0.1751209$, where $\alpha_0=-\langle\ln{|s|^2}\rangle =0.220051$ and 
$\beta_0=-\langle\frac{s^4}{|s|^4}=\frac{4}{\pi}-1=0.0170$.
From these we obtain $h(0)=0.73673$.

\begin{figure}
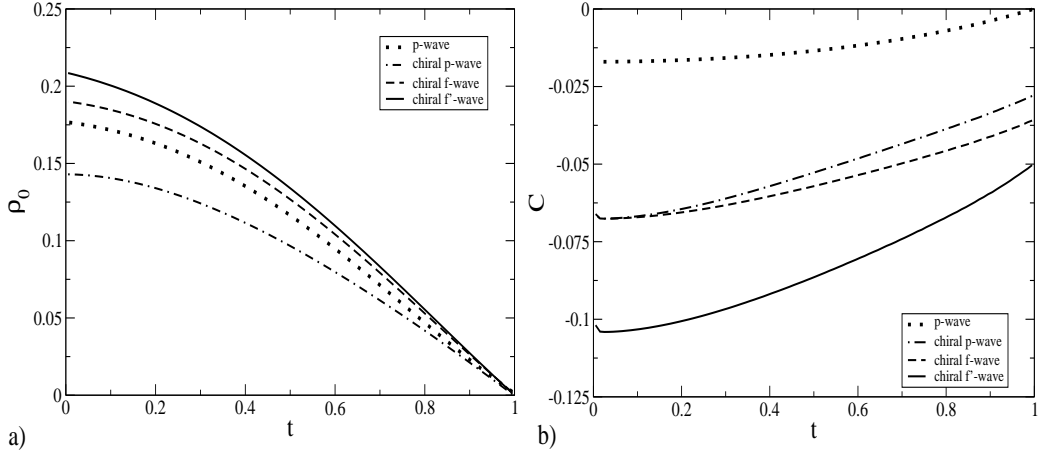

\begin{center}
\includegraphics[width=6.8cm,height=6cm,clip]{figures/campoa.eps}
\includegraphics[width=6.8cm,height=6cm,clip]{figures/ca.eps}
\caption{Normalised $H_{c2}(t)$ and $C(t)$ for $\vec{H}\parallel\vec{a}$ are shown in a) and b) respectively. Here solid, dashed, dashed-dotted and dotted lines are chiral f$^\prime$-wave, chiral f-wave, chiral p-wave and simple p-wave respectively.}
\label{campoa}
\end{center}
\end{figure}

Both $h(t)$ and $C(t)$ are evaluated numerically and we show them in Fig.~\ref{campoa} a) and b) respectively.

\subsection{B. Chiral p-wave SC: $\Delta(k)\sim \left(\frac{1}{\sqrt{2}}sgn(k_a)+\imath\sin{\chi_2}\right)$}

Now $H_{c2}(t)$ is determined by a similar set of equations as in sec. 1.B. In particular we find for
$t\rightarrow 1$ $C=-0.027735$ and $\rho = 0.212598 (\ln{t})$
while for $t\rightarrow 0$ $C=-0.067684$ and $\rho_0=0.139672$.
We obtain $h(0)=0.6566$. We show $h(t)$ and $C(t)$ in Fig.~\ref{campoa} a) and b) respectively.

\subsection{C. Chiral f-wave SC: $\hat{\Delta}(k)\sim \hat{d}s\cos{\chi_1}$}

Again we use a similar set of equations as those discussed in sec. 1.C, we find for $t\rightarrow 1$ $C=-0.0356236$ and $\rho = 0.2744495 (\ln{t})$ while for $t\rightarrow 0$
$C=0.066$ and $\rho_0=0.1920$ and $h(0)=0.6997$. Both $h(t)$ and $C(t)$ are evaluated numerically and shown in Fig.~\ref{campoa}
a) and b).

\subsection{D. Chiral f$^\prime$-wave SC: $\hat{\Delta}(k)\sim\hat{d}s\cos{\chi_2}$}

Now we find for $t\rightarrow 1$ $C=-0.05$ and
$\rho = -0.2910(\ln{t})$, while for $t\rightarrow 0$
$C=-0.1019$ and
$\rho_0=0.2090$.

We have shown again $h(t)$ and $C(t)$ in Fig.~\ref{campoa} a) and b) respectively.

Comparing these results with $H_{c2}(T)$ from (TMTSF)$_2$PF$_6$ and (TMTSF)$_2$ClO$_4$~\cite{KSD,Lee},
we can conclude  both $\vec{H}\parallel\vec{b^\prime}$ and $\vec{H}\parallel\vec{a}$ the chiral f$^\prime$-wave SC is most consistent with experimental data. In particular these states have relatively large $h(0)$ (see Table~\ref{table1}).On the other hand almost the same $H_{c2}(0)$ for $\vec{H}\parallel\vec{b^\prime}$ and
$\vec{H}\parallel\vec{a}$ has to be still accounted.

\begin{center}
\begin{table}[tbp]
\begin{center}
\caption{Summary of results. {\small Here $\rho_0(0)=\frac{\hat{v}^2eH_{c2}(0)}{2(2\pi T_c)^2}$ and $h(0)=\frac{H_{c2}(0)}{\frac{\partial H_{c2}(t)}{\partial t}\vert_t=1}$}}
\begin{tabular}{ccccccc}
& symmetry  & $C(0)$ & $C(1)$ & $-\frac{\partial \rho}{\partial t}\vert_t=1$ & $\rho_0(0)$ & $h(0)$\\
\hline
 & p-wave & -0.031 & 0 & 0.2377 & 0.1583 & 0.6659 \\
$\vec{H}\parallel b^\prime$ & chiral p-wave & -0.2247 & -0.3660 & 0.3838 & 0.2734 & 0.71324\\
 & chiral f$^\prime$-wave & -0.2247 & -0.3660 &  0.5181 & 0.3734 & 0.72073 \\
\hline
& p-wave & -0.017 & 0 & 0.2377 & 0.1751 & 0,7366 \\
$\vec{H}\parallel a$ & chiral p-wave & -0.066 & -0.028 & 0.2126 & 0.1396 & 0,6566 \\
 & chiral f-wave & -0.066 & -0.035 &  0.2744 & 0.1920 & 0.6997 \\
& chiral f$^\prime$-wave & -0.1019 & -0.05 &  0.2910 & 0.2090 & 0,7182 \\
\end{tabular}
\label{table1}
\end{center}
\end{table}
\end{center}

\section{Nodal lines in $\Delta(\vec{k})$}

We have seen that from the temperature dependence of $H_{c2}(T)$, we deduce the chiral f-wave and
chiral f$^\prime$ are the most favourable. They have nodal lines on the Fermi surface (i.e. the
$\chi_1-\chi_2$ plane), the chiral f-wave SC at $\chi_1=\pm \frac{\pi}{2}$, while chiral f$^\prime$-wave
SC at $\chi_2=\pm \frac{\pi}{2}$.

These nodal lines may be detected if the nuclear spin relaxation rate is measured in a magnetic
field rotated within the $b^\prime-c^*$ plane.

Following the standard procedure given in~\cite{Won} the quasiparticle density of states in the vortex
state for $T<<T_c$and $E=0$ is given by
\begin{equation}
N\left(0,\vec{H}\right)=\frac{2}{\pi^2}v^2\sqrt{eH}\left(1+\cos{\theta}^2\sin{\chi_{10}}^2\right)^\frac{1}{2}
\end{equation}
where $\chi_{10}$ is the position of the nodal line on the $\chi_{10}$ axis. So for the chiral f-wave SC
we find $\chi_{10}=\frac{\pi}{2}$ and $N\left(0,\vec{H}\right)$ exhibits the simple angular dependence.
On the other hand when nodal lines are on the $\chi_2$ axis, the $\theta$ dependence will be too small to see. Finally this gives

\begin{equation}
T_1^-1\left(\vec{H}\right)/T_{1N}^{-1}=\left(\frac{2}{\pi^2}\right)^2\vec{v}^2(eH)\left(1+\cos{\theta}^2\right)
\end{equation}
for the chiral f-wave SC.

\begin{figure}[t]
\begin{center}
\includegraphics[width=8cm]{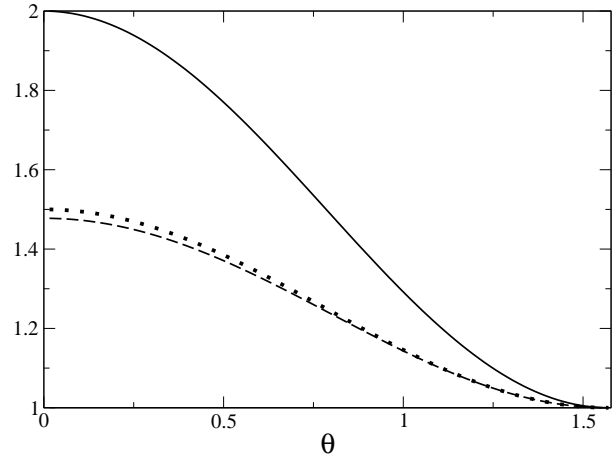}
\caption{The angle dependent nuclear spin relaxation rate for a few nodal superconductors is shown. {\small (Chiral f-wave, chiral h-wave and chiral p-wave are represented in solid, dashed and dotted lines.)}}
\end{center}
\label{t}
\end{figure}

We show the $\theta$ dependence of $T_1^{-1}$ in Fig.~\ref{t} for a few candidates. The chiral f-wave SC has the strongest $\theta$ dependence (solid line) while the chiral h-wave SC (dashed line) and the chiral p-wave SC (dotted line) have a similar $\theta$ dependence.

\section{Concluding remarks}

We have computes the upper critical field of Bechgaard salts for a variety of model superconductors
with the standard microscopic theory. We find: a) Assuming all these superconductors have the same $T_c$, the chiral f$^\prime$-wave SC
($\vec{\Delta}(k)\sim \left( \frac{1}{\sqrt{2}}sgn(K_a)+\imath \sin{\chi_2} \right) \cos{\chi_2}$)
appears to be the most favourable with largest $H_{c2}$'s for both $\vec{H}\parallel b^\prime$ and
$\vec{H}\parallel a$; b) however, non of these states exhibit the quasi T linear dependence of $H_{c2}(T)$ as observed in
~\cite{KSD}; c) Also the present theory predicts $H_{c2}(0) \sim \left( v\,v_c\right)^{-1}$ and $\left(v_b v_c\right)^{-1}$
for  $\vec{H}\parallel b^\prime$ and $\vec{H}\parallel a$ respectively. This means $H_{c2}(0)$ for $\vec{H}\parallel a$ is about 5 time larger than the one for
$\vec{H}\parallel b^\prime$ contrary to observation; d) from $H_{c2}(0) \sim 5$T and $T_c=1.5$K we can extract $V^2=\sqrt{v\,v_c}\sim 1.5*10^4$cmás$^{-1}$, consistent with the known values of $v$, $v_c$.

\acknowledgements
We thank Stuart Brown and Paul Chaikin for useful discussion on possible detection of the nodal structure of $\Delta(\vec{k})s$ in Bechgaard salts through NMR. We have benefited from discussion with Stephan Haas and David Parker.
ADF acknowledges gratefully the financial support of Ministerio de Educaci\'on y Ciencia (Spain) (AP2003-1383) and Jaime Ferrer and Paco Guinea for useful discussion.

\end{document}